\documentclass[12pt]{article} %\input epsf.tex
\pdfoutput=1
\usepackage[utf8]{inputenc}
\usepackage[T1]{fontenc}
\usepackage{lmodern, amsmath, amssymb, slashed, graphicx, comment,adjustbox}
\usepackage[dvipsnames]{xcolor}
\usepackage[numbers, elide, compress]{natbib}
\usepackage[colorlinks=true, citecolor=blue, urlcolor=blue, linkcolor=blue, breaklinks=true, pdfpagelabels=false]{hyperref}
\usepackage{caption}
\usepackage{subcaption}

\makeatletter\g@addto@macro\bfseries{\boldmath}\makeatother

\def\figureautorefname~#1\null{Fig.\,#1\null}

\def\equationautorefname~#1\null{Eq.\,(#1)\null}

\numberwithin{equation}{section}

\newcommand{\inab}{\,{\rm ab}^{-1}}
\newcommand{\infb}{\,{\rm fb}^{-1}}

\newcommand{\Afb}[1]{A^{0,#1}_{\rm FB}}

\newcommand{\bpm}{\begin{pmatrix}}
\newcommand{\epm}{\end{pmatrix}}

\usepackage{pifont}
\begin{document}

\begin{flushright}
%xxxx
\end{flushright}

\vspace*{1.5cm}

\begin{center}

{\Large\bf
%A New Look in the Beautiful Mirror from the W-boson Mass Measurement
Accommodating the CDF W-boson Mass Measurement in the Beautiful Mirror Model
\par}
\vspace{9mm}

{\bf Shengdu~Chai$\,^{a,c}$},~~~{\bf Jiayin~Gu$\,^{a,b}$},~~~{\bf Lian-Tao Wang$\,^{c}$}\\ [4mm]
{\small\it
$^a$ Department of Physics and Center for Field Theory and Particle Physics, \\ Fudan University, Shanghai 200438, China \\[2mm]
$^b$ Key Laboratory of Nuclear Physics and Ion-beam Application (MOE), \\ Fudan University, Shanghai 200433, China  \\[2mm]
$^c$ Department of Physics and Enrico Fermi Institute, University of Chicago, Chicago, IL 60637
\par}
\vspace{.5cm}
\centerline{\tt \small sdchai19@fudan.edu.cn,  jiayin\_gu@fudan.edu.cn, liantaow@uchicago.edu}

\end{center}

\vspace{1cm}

\begin{abstract}
The W-boson mass measurement recently reported by the CDF II experiment exhibits a significant deviation from both the Standard Model prediction and previous measurements.  There is also a long-standing deviation between the Standard Model prediction of the forward-backward asymmetry of the bottom quark ($\Afb{b}$) and its measurement at the LEP experiment.  The Beautiful Mirror model, proposed to resolve the $\Afb{b}$ discrepancy, introduces vector-like quarks that modify the W-boson mass at one-loop level.  In this study, we find an interesting region in the model parameter space that could potentially explain both discrepancies, which puts the new quarks in the multi-TeV region.  This region is mostly consistent with current LHC bounds from direct searches and Higgs coupling measurements, but will be thoroughly probed at the High Luminosity LHC. As such, the Beautiful Mirror model as an explanation of the $m_W$ and  $\Afb{b}$ discrepancies could be confirmed or falsified in the near future.  
\end{abstract}

\newpage
{\small 
\tableofcontents}

\setcounter{footnote}{0}
%\pagebreak
%%%%%%%%%%%%%%%%%%%%%%%%%%%%%%%%%%%%%%%%%%%%%%%%%%%%%%%%%%

%%%%%%%%%%%%%%%%%%%%%%%%%%%%%%%%%%%%%%%%%%%%%%%%%%%%%%%%%%
\section{Introduction}
%%%%%%%%%%%%%%%%%%%%%

Recently, the CDF II experiment reported a new measurement of the $W$ boson mass~\cite{CDF:2022hxs},
\begin{equation}
m_W^{\rm CDF-II} = 80433.5\pm 9.4~{\rm MeV}\,,
\end{equation}
which exhibits a $7\sigma$ deviation from the Standard Model (SM) prediction, $m^{\rm SM}_W = 80357\pm 6~{\rm MeV}$, and a $4\sigma$ deviation from the %previous 
PDG world average value~\cite{ParticleDataGroup:2022pth}, $m^{\rm PDG}_W =  80377\pm 12~{\rm MeV}$.
The possible new physics explanations of this discrepancy have been extensively studied afterwards~\cite{Lu:2022bgw, DAlise:2022ypp, Zhu:2022tpr, Fan:2022dck, Strumia:2022qkt, Athron:2022qpo, deBlas:2022hdk, Tang:2022pxh, Du:2022pbp, Campagnari:2022vzx, Cacciapaglia:2022xih, Blennow:2022yfm, Sakurai:2022hwh, Fan:2022yly, Zhu:2022scj, Arias-Aragon:2022ats, Liu:2022jdq, Paul:2022dds, Babu:2022pdn, Gu:2022htv, DiLuzio:2022xns, Bagnaschi:2022whn, Heckman:2022the,  
Bahl:2022xzi, Song:2022xts, Asadi:2022xiy, Athron:2022isz, Heo:2022dey, Crivellin:2022fdf, Endo:2022kiw, Du:2022brr, Cheung:2022zsb, DiLuzio:2022ziu, Balkin:2022glu, Biekotter:2022abc, Krasnikov:2022xsi, Zheng:2022irz, Ahn:2022xax, Han:2022juu,Kawamura:2022uft, Ghoshal:2022vzo, FileviezPerez:2022lxp, Nagao:2022oin, Kanemura:2022ahw, Mondal:2022xdy, Wilson:2022gma, Zhang:2022nnh, Cirigliano:2022qdm, Borah:2022obi, Chowdhury:2022moc, Arcadi:2022dmt, Popov:2022ldh, Carpenter:2022oyg, Bhaskar:2022vgk, Ghorbani:2022vtv, Du:2022fqv, Zeng:2022lkk, Batra:2022org, Borah:2022zim, Cao:2022mif, Baek:2022agi, Heeck:2022fvl, Addazi:2022fbj, Cheng:2022aau, Almeida:2022lcs, Lee:2022gyf, Cai:2022cti, Benbrik:2022dja, Yang:2022qgs, Batra:2022pej, Tan:2022bip, Abouabid:2022lpg, Chen:2022ocr, Gupta:2022lrt, Basiouris:2022wei, Wang:2022dte, Botella:2022rte, Kim:2022xuo, Kim:2022hvh, Barman:2022qix, He:2022zjz, Li:2022gwc, Dcruz:2022dao, Thomas:2022gib, Isaacson:2022rts, Appelquist:2022qgl, Gao:2022wxk, Evans:2022dgq, Kim:2022zhj, Chowdhury:2022dps, Senjanovic:2022zwy, Lazarides:2022spe, Ghosh:2022zqs, Miralles:2022jnv, Liu:2022vgo, Asai:2022uix, Ma:2022emu, Kawamura:2022fhm, Chen:2022lwc, Allanach:2022bik, Afonin:2022cbi, Hill:2022ayj, Xue:2022mde, Rizzo:2022jti, VanLoi:2022eir, YaserAyazi:2022tbn, Chakrabarty:2022voz, CentellesChulia:2022vpz, Nagao:2022dgl, Frandsen:2022xsz, Bahl:2022gqg, Arora:2022uof, Abdallah:2022shy, Batra:2022arl, Benakli:2022gjn, Barger:2022wih, Domingo:2022pde, Rodriguez:2022hsj, Liu:2022zie, Diaz:2022vdo, Lin:2022khg, Chung:2022avf, Butterworth:2022dkt, Bandyopadhyay:2022bgx}. 
These explanations can be generally divided into two categories. One could introduce new physics that modify the $W$ mass at the tree level.  This puts the new particle masses at or above the multi-TeV range which is beyond the reach of current or  future LHC searches, but the new physics scenarios are limited to only a few possibilities, such as triplet scalar models (see {\it e.g.} Ref.~\cite{Kanemura:2022ahw}).  A larger class of models could modify the $W$ mass at the one-loop level, which leads to much richer phenomenological implications.  In this case, larger couplings or smaller new-particle masses are usually required to generate a large enough modification to the $W$ mass.  This is generally in tension with direct search bounds, or other precision measurements if the new physics contributes at the tree level.

On the other hand, there is a long-standing $2.5\sigma$ discrepancy in the forward backward asymmetry of the bottom quark ($\Afb{b}$) between the SM prediction and the measured value at the LEP experiment~\cite{ALEPH:2005ab}.  Assuming this discrepancy is due to effects of beyond-Standard-Model (BSM) physics, it could be explained by introducing new exotic quarks that mix with the bottom quark, first proposed in Ref.~\cite{Choudhury:2001hs} with the name ``Beautiful Mirror'' (BM).  The new quarks contribute to the propagators of the electroweak gauge bosons and generate a non-zero $T$ parameter, and thus modify the $W$-boson mass.  
This offers an intriguing possibility that the two measurement discrepancies could come from the same physics origin.

In this study, we further investigate the possibility that the BM Model gives rise to the measured discrepancies (from SM) for both $\Afb{b}$ at LEP and $m_W$ at CDF II.  By performing a global fit of electroweak precision measurements, we find the preferred region of the parameter space.  As shown later, it indeed provides a good fit to both measurements, significantly reducing the tension by only introducing a few model parameters in addition to the SM ones.  We also study other phenomenological implications of this model, including the direct searches for exotic quarks and the precision Higgs measurements at the LHC.  Roughly speaking, the parameter space preferred by the $\Afb{b}$ and CDF-II $m_W$ measurements is consistent with current LHC measurements but can be thoroughly probed by future LHC runs, especially the HL-LHC.  Therefore, the possibility that both the $\Afb{b}$ and CDF-II $m_W$ discrepancies are explained by the BM model can be confirmed or falsified in the near future.  

The rest of this paper is organized as follows:  In \autoref{sec:bm} we review the BM model and work out its modifications to the relevant observables.  In \autoref{sec:ew}, we perform an electroweak global fit to find the preferred region in the model parameter space.  In \autoref{sec:lhc}, we summarize the current status of the relevant direct search bounds and Higgs coupling measurements as well as their future projections at the HL-LHC.  The preferred regions of the model parameter space from various measurements are presented in \autoref{sec:result}.  Finally, we conclude in \autoref{sec:con}.

%%%%%%%%%%%%%%%%%%%%%%%%%%%%%%%%%%%%%%%%%%%%%%%%%%%%%%%%%%
\section{The Beautiful Mirror model}
\label{sec:bm}
%%%%%%%%%%%%%%%%%%%%%%

The BM model was proposed in Ref~\cite{Choudhury:2001hs} as a new-physics explanation of the $\Afb{b}$ measurement at LEP~\cite{ALEPH:2005ab}, which deviates significantly ($\sim 2.5\sigma$) from the SM prediction.   New vector-like exotic quarks are introduced, which mix with the bottom quark and modifies the $Zb\bar{b}$ couplings.  Defining $\delta g_{Lb}$ and $\delta g_{Rb}$ as
\begin{equation}
\mathcal{L} = -\frac{g}{c_W} Z_\mu \left[ \bar{b}_L \gamma^\mu (-\frac{1}{2} +\frac{1}{3} s^2_W + \delta  g_{Lb} ) b_L  +  \bar{b}_R \gamma^\mu (\frac{1}{3} s^2_W + \delta  g_{Rb} ) b_R  \right] + \ldots \,,
\end{equation}      
where $s_W \equiv \sin\theta_W$, $c_W \equiv \cos\theta_W$ and $\theta_W$ is the weak mixing angle, positive values are preferred for both $\delta g_{Lb}$ and $\delta g_{Rb}$, as shown later in \autoref{sec:ew}.  This can be achieved by introducing two vector-like quarks,   
\begin{align}
\Psi_{L,R} =&~ \bpm B \\ X \epm   \sim (3,2)_{-5/6} \,, \nonumber\\
\hat{B}_{L,R}  \sim & ~~  (3,1)_{-1/3} \,,  \label{eq:BM}
\end{align}
where the numbers in the bracket denote representations under $SU(3)_c$ and $SU(2)_W$, respectively, and the subscript denotes $U(1)_Y$ hypercharge.  The mass and relevant interaction terms in the Lagrangian are given by 
\begin{equation}
-\mathcal{L} \supset M_1 \bar{\Psi}_L \Psi_R + M_2 \bar{\hat{B}}_L \hat{B}_R 
+y_1 \bar{Q}_L H b_R+ y_L \bar{Q}_L H \hat{B}_R + y_R \bar{\Psi}_L \tilde{H} b_R +\mbox{h.c.} \,.  \label{eq:Lbm}
\end{equation}
After the electroweak symmetry breaking, %(assuming $H=\frac{1}{\sqrt{2}}(0,v+h)^T$), 
the vacuum expectation value (VEV) of the Higgs field generates mixings between the new quarks and the SM bottom quark, and modifies the $Zb\bar{b}$ couplings as
\begin{equation}
\delta g_{Lb} = \frac{y^2_L v^2}{4M^2_2}  \,, ~~~~~   \delta g_{Rb} =  \frac{y^2_R v^2}{4M^2_1}  \,, \label{eq:y123}
\end{equation}
where $v=246\,$GeV. They can be obtained by either diagonalizing the mass matrix or using effective-field-theory (EFT) methods,\footnote{The sum rules listed in Ref.~\cite{Gu:2020thj} are also particularly convenient for the calculation in the EFT approach.} and are both positive as desired. 
Contributions to the gauge boson propagators are generated at the one loop, which modifies the $S$ and $T$ parameters~\cite{Peskin:1991sw}.  A direct computation gives 
\begin{align}
    S\approx&~ \frac{2}{9\pi}\left[-2\delta g_{Rb}{\left(\log{\left(\frac{y^2_1v^2}{2M^2_1}\right)} +7\right)}+\delta g_{Lb}{\left(4 \log{\left(\frac{y^2_1v^2}{2M^2_2}\right)}+2 \right)}\right] \, ,\label{eq:Ssimp}  \\
T \approx&~ \frac{3}{16\pi^2 \alpha v^2} \left[ \frac{16}{3} \delta g_{Rb}^2 M^2_1 + 4 \delta g_{Lb}^2 M^2_2 - 4\delta g_{Lb} \frac{M^2_2 \, m^2_{\rm top}}{M^2_2-m^2_{\rm top}} \log{\left(\frac{M^2_2}{m^2_{\rm top}}\right)}    \right] \,,  \label{eq:Tsimp}
\end{align}
where we have omitted terms suppressed by the small bottom mass.  Note that the $S$-parameter is generally small since we have only introduced vector-like quarks.  On the other hand, there is a sizable contribution to the $T$-parameter. 

The BM model also modifies the Higgs couplings, and can therefore also be probed by precision Higgs measurements.  At the tree level, only the $hb\bar{b}$ coupling is modified.  A large set of Higgs couplings will be modified at the one-loop level.  However, we expect the $hgg$ and $h\gamma\gamma$ couplings to be the most relevant ones since they are relatively well probed and their leading SM contributions are also at the one-loop level.  For simplicity, we focus on the modifications of the $hb\bar{b}$, $hgg$ and $h\gamma\gamma$ couplings, and compute them in the $m^2_h \ll M_{1,2}$ limit. 
They can be parameterized by the familiar ``kappa'' parameterization,
\begin{equation}
    \frac{\Gamma(h\to b\bar{b})}{\Gamma_{\rm SM}(h\to b\bar{b})} \equiv (1+\delta \kappa_b)^2 \,, \hspace{0.5cm}  
    \frac{\Gamma(h\to gg)}{\Gamma_{\rm SM}(h\to gg)}\equiv (1+\delta \kappa_g)^2 \,, \hspace{0.5cm} 
    \frac{\Gamma(h\to \gamma \gamma)}{\Gamma_{\rm SM}(h\to \gamma \gamma)}\equiv (1+\delta \kappa_{\gamma})^2 \,.
\end{equation}
In particular, in the BM model they can be connected to $\delta g_{Lb}$ and $\delta g_{Rb}$ as (see also \cite{Batell:2012ca})
\begin{equation}
\delta \kappa_b \simeq -2 \,(\delta g_{Lb}+\delta g_{Rb}) \,, \hspace{0.5cm}
\delta \kappa_g \simeq 1.937 \,(\delta g_{Lb}+\delta g_{Rb}) \,,  \hspace{0.5cm}
\delta \kappa_\gamma \simeq  -0.137\, (\delta g_{Lb}+\delta g_{Rb}) \,. \label{eq:deltakappa}
\end{equation}

%%%%%%%%%%%%%%%%%%%%%%%%%%%%%%%%%%%%%%%%%%%%%%%%%%%%%%%%%%
\section{Electroweak global fit}
\label{sec:ew}
%%%%%%%%%%%%%%%%%%%%%%

\begin{figure}%[h!]
    \centering
    \includegraphics[width=0.47\textwidth]{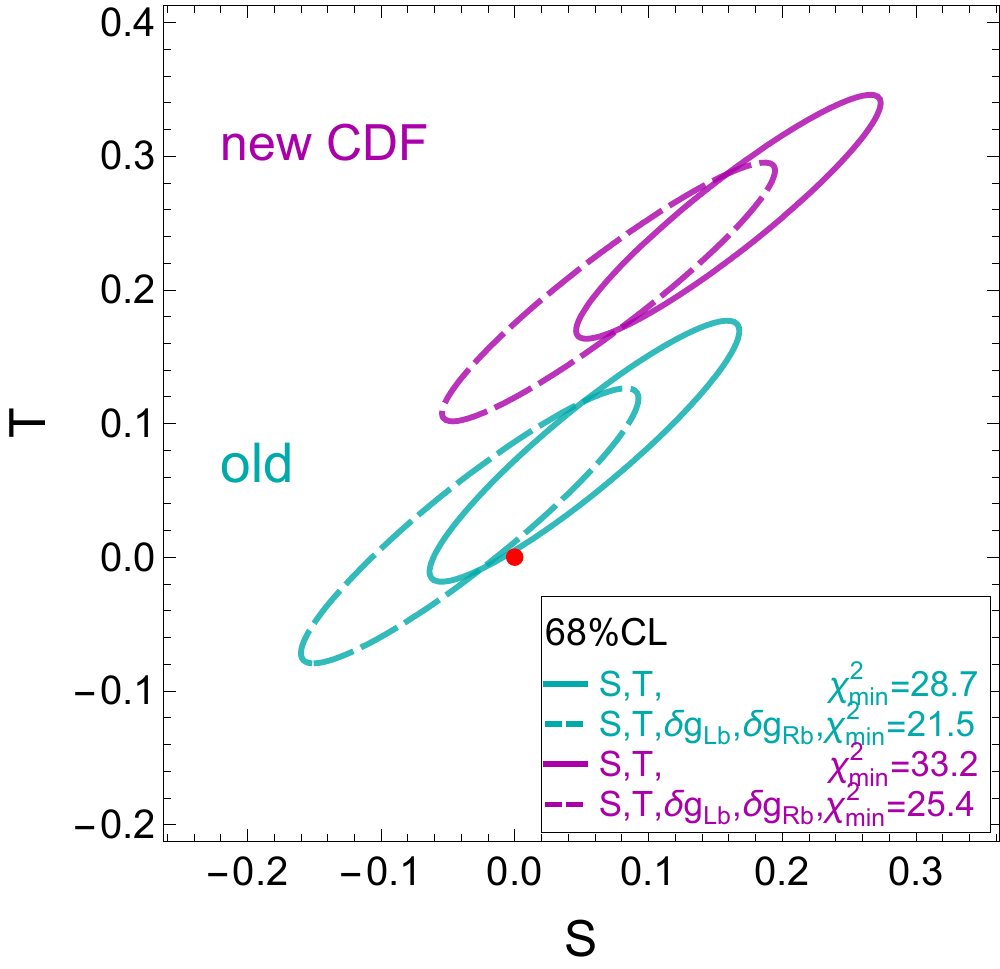}
    \hspace{0.3cm}
    \includegraphics[width=0.47\textwidth]{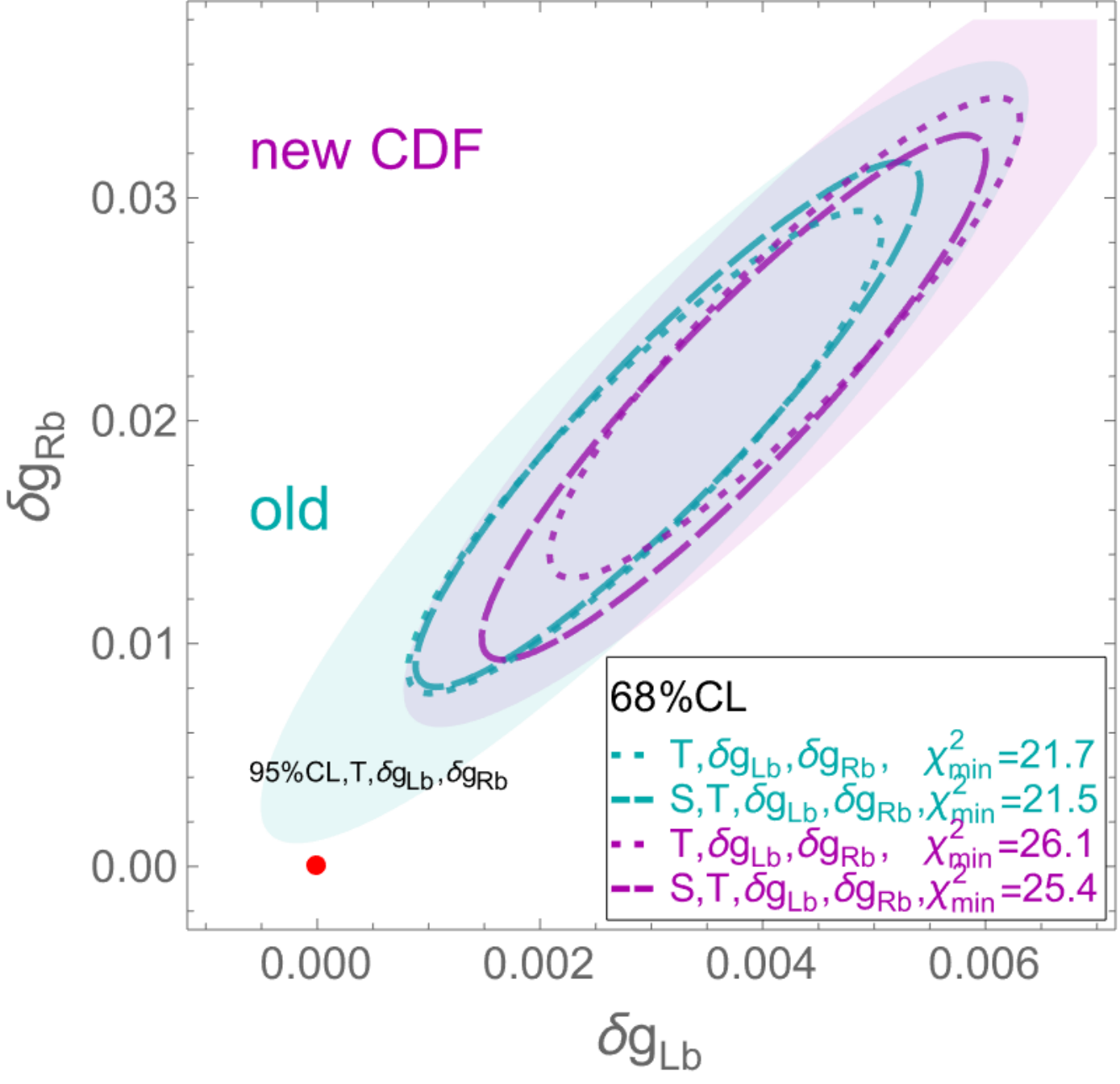}
    \caption{Preferred regions in the $S-T$ and $\delta g_{Lb}-\delta g_{Rb}$ planes from a 2-parameter fit with $S$ and $T$ (solid contours), a 4-parameter fit with $S,T,\delta g_{Lb},\delta g_{Rb}$ (dashed contours), {and a 3-parameter fit with $T,\delta g_{Lb},\delta g_{Rb}$ (fixing $S=0$, dotted contours)} with the current EW measurements.  Two different $m_W$ scenarios are considered, which are the “new CDF” measurement and the “old” PDG $m_{W}$ measurement before the CDF one.  The red point is the SM prediction. All contours correspond to 68\% confidence level, {except for the shaded regions on the right panel which correspond to the 95\%CL region of the 3-parameter fit with $T,\delta g_{Lb},\delta g_{Rb}$.}}
    \label{fig:EWfit}
\end{figure}

We perform a  global fit to Electroweak observables to determine the preferred region in the parameter space of the BM model.  Following Ref.~\cite{Gori:2015nqa}, the fit is done in the ``SM$+S,T,\delta g_{Lb},\delta g_{Rb}$'' framework, 
where the SM is augmented by the four free parameters $S$, $T$, $\delta g_{Lb}$ and $\delta g_{Rb}$, which contain the leading contributions of the BM model.  The results could then be easily translated into constraints on the model parameters.  
Our analysis mainly follows Refs.~\cite{Efrati:2015eaa, Falkowski:2014tna, Gu:2022htv}, and is a slightly simplified version of the ones in Refs.~\cite{Batell:2012ca, Gori:2015nqa}.  In particular, we have checked that our results are in good agreement with the one in Ref.~\cite{Gori:2015nqa}. 
We choose the $\{ \alpha,m_{Z}, G_F\} $ input scheme and fix the values of the input parameters as~\cite{ParticleDataGroup:2022pth}
\begin{equation}
    \alpha = 1/127.940 \,,~~~~ m_Z= 91.1876\, {\rm GeV}\,,~~~~G_F=1.1663787\times10^{-5}\, {\rm GeV}\,. \label{eq:input3}
\end{equation}
We include the $W$ and $Z$ pole measurements, which are 
\begin{equation}    \Gamma_{Z},~~~\sigma_{had},~~~R_{f},~~~A_{FB}^{0,f},~~~A_{f},~~~A_{e/\tau}^{pol} \,, \nonumber \\
\end{equation}
\begin{equation}
     m_W,~~~\Gamma_W,~~~\rm Br(W\to e\nu),~~~\rm Br(W\to \mu\nu),~~~\rm Br(W\to\tau \nu) \,,
\end{equation}
where $f=e,u,\tau,b,c$, and $A_{e/\tau}^{pol}$ is $A_e$ and $A_\tau$ measured using final-state tau polarizations at LEP.   
For the Z-pole measurements, we use the results in Ref.~\cite{ALEPH:2005ab}. For the W branching ratios measurements, we take the results from Ref.~\cite{Schael:2013ita}. The measurement of $\Gamma_W$ is taken from Ref.~\cite{ParticleDataGroup:2022pth} . For the $W$ boson mass $m_W$, we consider two different measurements, one is the ``old'' world average measurement from Ref.~\cite{ParticleDataGroup:2022pth}, the other is the ``new CDF'' measurement from Ref.~\cite{CDF:2022hxs}.  
Our results are shown in \autoref{tab:STLR} in terms of the 1-sigma bounds of $S,T,\delta g_{Lb},\delta g_{Rb}$ and their correlations.  We further illustrate the  results in \autoref{fig:EWfit} in terms of the $68\%$ confidence-level (CL) regions in the $S$-$T$ and $\delta g_{Lb}$-$\delta g_{Rb}$ planes.  For comparison, the results in the ${\rm SM}+S,T$ framework (with $\delta g_{Lb}$, $\delta g_{Rb}$ fixed to zero) {and the ones in the ${\rm SM}+T,\delta g_{Lb},\delta g_{Rb}$ framework (with $S=0$)} are also shown.  
{The latter scenario is closer to the case in the BM model, which generally has a very small contribution to the $S$ parameter, as mentioned above.}

\begin{table}%[h]
\centering
\begin{tabular}{|c||c|cccc|} \hline
 & $1\sigma$ bound &   \multicolumn{4}{c|}{correlation}      \\ %\cline{2-13}
 &  &    $~S~$  &  $~T~$  & $\delta g_{Lb}$ & $\delta g_{Rb}$    \\ \hline \hline
  &\multicolumn{5}{|c|}{old (PDG)}  \\  \hline
$S$    &   $-0.034\pm 0.084$     &    1 &  &   &    \\
$T$    &   $0.023\pm0.068$     &  0.926   & 1  &   &      \\
$\delta g_{Lb}$    &   $0.0031\pm0.0015$     &  $-0.345$   & $-0.237$ & 1  &  \\
$\delta g_{Rb}$    &   $0.020\pm 0.0078$     &   $-0.394$  & $-0.299$ & 0.917  & 1  \\\hline\hline
& \multicolumn{5}{|c|}{new CDF}  \\  \hline
$S$    &   $0.070\pm 0.082$     &    1 &  &   &    \\
$T$    &   $0.198\pm0.064$     &  0.938   & 1  &   &      \\
$\delta g_{Lb}$    &   $0.0037\pm0.0015$     &  $-0.359$   & $-0.271$ & 1  &  \\
$\delta g_{Rb}$    &   $0.021\pm 0.0078$     &   $-0.403$  & $-0.326$ & 0.918  & 1  \\\hline
\end{tabular}
\caption{Best fit values $\pm 1 \sigma$ of and correlations among $S,T,\delta g_{Lb},\delta g_{Rb}$ from EW global fit with the two different $m_W$ scenarios. 
}
\label{tab:STLR}
\end{table}

It is clear in \autoref{fig:EWfit} that, with only the $S$ and $T$ parameters, the new CDF W-mass measurement prefers a positive shift in both $S$ and $T$.  This is in agreement with many of the earlier EW-fit results, such as Refs.~\cite{Lu:2022bgw, Strumia:2022qkt, deBlas:2022hdk, Fan:2022yly, Gu:2022htv, Bagnaschi:2022whn}.  The inclusion of $\delta g_{Lb}$, $\delta g_{Rb}$ brings down the overall $\chi^2_{\rm min}$ due to the better agreements in $\Afb{b}$, and also introduces a negative shift in $S$ and $T$ (as also observed in Ref.~\cite{Gori:2015nqa}).  As a result, a zero $S$ is within the 68\%CL even with the new CDF $m_W$ measurement.  On the other hand, as shown in \autoref{fig:EWfit}, the 68\%\,CL preferred region in the $\delta g_{Lb}$-$\delta g_{Rb}$ plane changes only slightly under a shift in $m_W$, and the SM point $(0,0)$ is clearly at the outside of it. 
{However, the difference is notably larger for the $S=0$ case, with the preferred regions of $\delta g_{Lb}$ and $\delta g_{Rb}$ both shifted to larger values with the new CDF $m_W$ measurement.  This has important implications in the BM model, as we will discuss below.} 
\begin{figure}[t]
    \centering
    \includegraphics[width=0.45\textwidth]{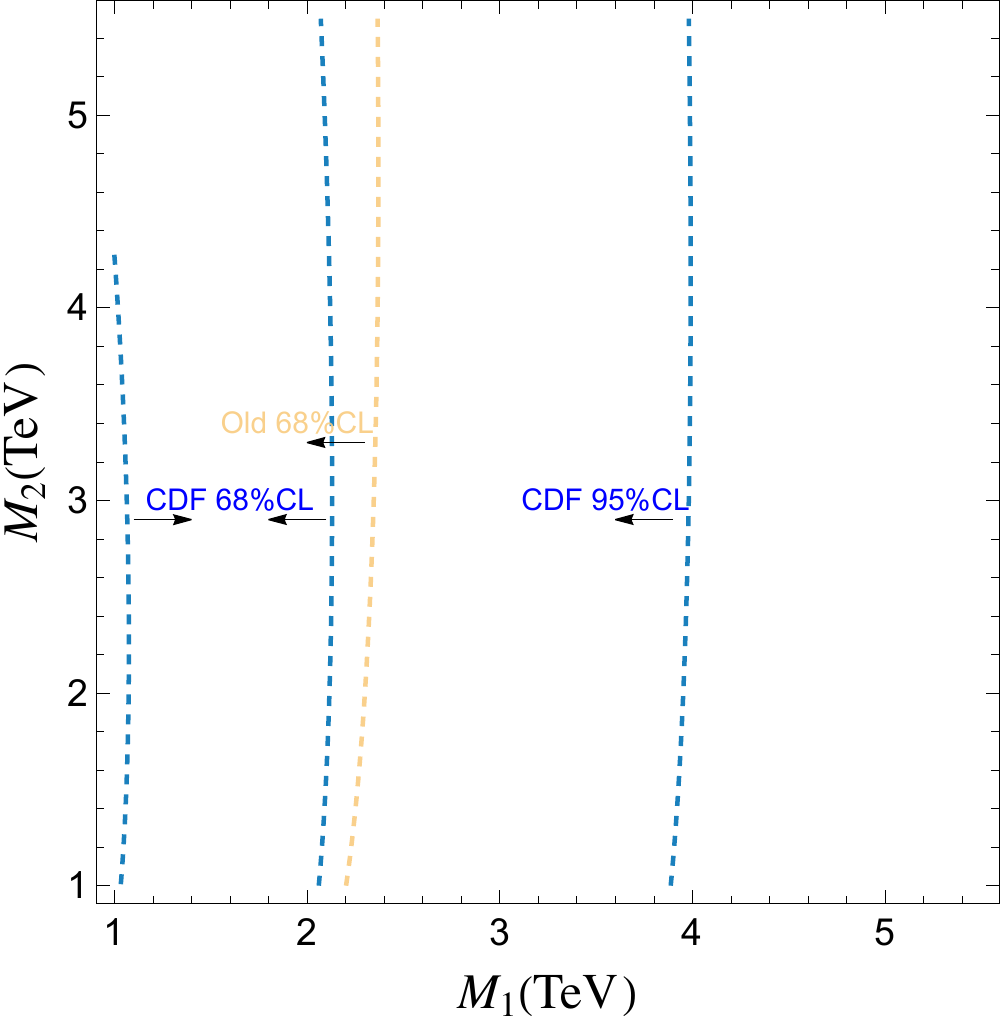} \hspace{0.3cm}
    \includegraphics[width=0.47\textwidth]{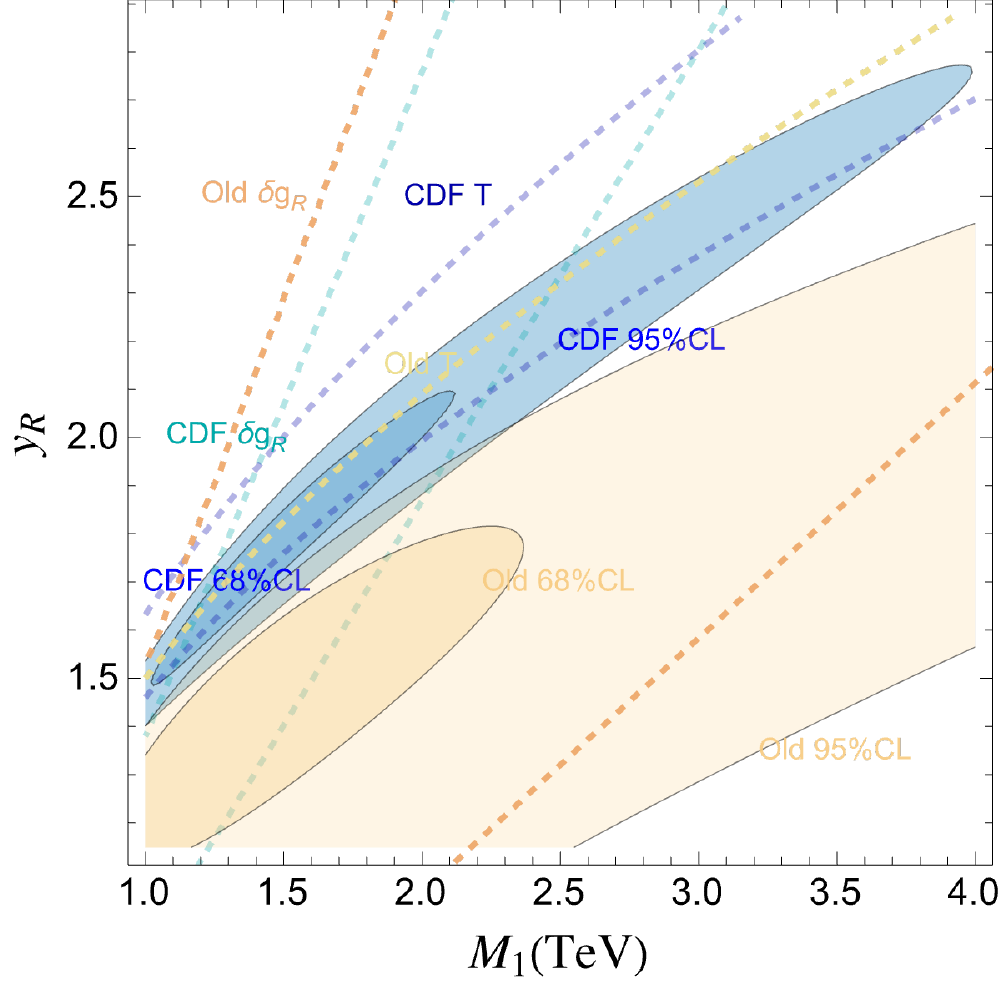}
    \caption{
    {\bf Left:} 68\% and 95\%\,CL preferred regions in the $M_1-M_2$ plane from the EW fit with the new CDF $m_W$ {(blue) and the old PDG $m_W$ (orange).  The arrows point to the more preferred regions.  For the ``old'' case, the 95\%CL contour is at $M_1 \simeq 10$\,TeV.}
    %, with $\delta g_{Lb}$ and $\delta g_{Rb}$ fixed to the best fitted values ($\delta g_{Lb}=0.0037,\delta g_{Rb}=0.021 $). 
    {\bf Right:} Preferred regions of parameter space in the $M_1-y_R$ plane, with $M_2$ fixed at $4\,{\rm TeV}$ and $y_L$ marginalized. 
    The blue {(orange)} areas are the 68\% and 95\%\,CL regions from the EW global fit with the new CDF $m_W$ {(old PDG $m_W$)}, while the cyan, blue {(orange, yellow)} dashed lines correspond to $68\%$\,CL bounds from the $\delta g_{Rb}$, $T$-parameter constraint alone, respectively.
    }
    \label{fig:bmfit1}
\end{figure}
In \autoref{fig:bmfit1}, we project the EW global-fit results (with the new CDF $m_W$) onto the parameter space of the BM model.  The relevant parameters are the two Yukawa couplings $y_R$, $y_L$, and the two masses, $M_1$ and $M_2$.  
A crucial observation is that while $\delta g_{Lb}$, $\delta g_{Rb}$ and $T$ all scales as $1/M^2_{1,2}$ at the leading order as they correspond to dimension-6 operators, their dependencies on the Yukawa couplings are different, with $\delta g_{Lb}, \delta g_{Rb} \sim y^2$ and $T \sim y^4$.  This difference can be easily understood from the fact that $\delta g_{Lb}$, $\delta g_{Rb}$ are generated at the tree level with 2 insertions of Yukawa couplings, while $T$ is generated by a fermion loop with 4 insertions of Yukawa couplings (as $\mathcal{O}_{T} = \frac{1}{2}(H^\dagger \overleftrightarrow{D_\mu} H)^2$ contain 4 external Higgs legs).
Therefore, the global fit could in principle separately constrain the Yukawa couplings and the mass terms.  This is illustrated on the left panel of \autoref{fig:bmfit1}, where we %fix $\delta g_{Lb}$ and $\delta g_{Rb}$ to their best-fit value and scan over the parameter space of $(M_1, \,M_2)$, in which case the bound is mostly provided by the $T$ parameter.  
{marginalize over $\delta g_{Lb}$ and $\delta g_{Rb}$ and project the bounds from the ``SM$+S,T,\delta g_{Lb},\delta g_{Rb}$'' fit on the $(M_1, \,M_2)$ plane.  Results with the ``new CDF'' $m_W$ (in blue) and the ``old'' PDG $m_W$ (in orange) are shown in comparison. Note that, for fixed $\delta g_{Lb}$ and $\delta g_{Rb}$, the bound is mostly provided by the $T$ parameter.} 
The result clearly shows that the EW measurements are mainly sensitive to $M_1$, which is expected from \autoref{eq:Tsimp} since $\delta g_{Rb} \gg \delta g_{Lb}$, {and the upper bound on the $T$ parameter also puts an upper bound on $m_W$.}  
{On the other hand, it is peculiar that with the new CDF $m_W$, the upper bound on $M_1$ is stronger despite that it allows for a larger value of $T$.  This is because, as shown in \autoref{fig:EWfit}, the preferred value of $\delta g_{Rb}$ is also larger with the new CDF $m_W$, especially for $S=0$ which is approximately the case in the BM model.  It is clear from \autoref{eq:Tsimp} that for a fixed $T$ this would decrease the value of $M_1$. 
In particular, for the ``old'' measurements, the 95\%CL contour almost reaches the point $\delta g_{Lb}=\delta g_{Rb}=0$, at which the $T$ parameter would not provide an upper bound on $M_1$.  Indeed,  we observe a much larger 95\%CL upper bound for $M_1$ with old measurements, which is at around 10\,TeV.}
%
%As such, 
On the right panel of \autoref{fig:bmfit1} we fix $M_2$ to a relatively large value, $4\,$TeV, and marginalize over $y_L$ to find the preferred region in the $(M_1, \,y_R)$ plane from the ``SM$+S,T,\delta g_{Lb},\delta g_{Rb}$'' fit.  This is our main result, which has important implications.  First, it is clear that $y_R$ and $M_1$ can be separately constrained due to the interplay between $\delta g_{Rb}$ and $T$.  
To illustrate this, we also show the 68\%\,CL bounds from the $\delta g_{Rb}$ alone %(cyan dashed lines) 
and $T$ alone, %(blue dashed lines), 
which clearly constrain different combinations of $y_R$ and $M_1$.   
As a result, {with the new CDF $m_W$ measurement (blue regions),} the global fit puts an upper bound on $M_1$, with $M_1\lesssim 4\,$TeV within a 95\%\,CL.  Crucially, this bound is consistent with the current LHC limits but could be relevant for future LHC probes, as we will discuss in the next section. 
{On the other hand, with the old $m_W$, a larger $M_1$ is allowed, as mentioned above.} 
The corresponding $y_R$ is of order one {in both cases}, which ensures the perturbativity of the model.  

%%%%%%%%%%%%%%%%%%%%%%%%%%%%%%%%%%%%%%%%%%%%%%%%%%%%%%%%%%
\section{LHC probes}
\label{sec:lhc}
%%%%%%%%%%%%%%%%%%%%%%

With the BM model being a potential explanation of both the $\Afb{b}$ and the $m_W$ discrepancies, it is crucial to look for additional signals at the LHC that could confirm (or rule out) this possibility.  The signals could come from either the direct searches of heavy exotic quarks or the precision Higgs measurements.  They are discussed separately in this section.  

%%%%%%%%%%%%%%%%%%%%%%
\subsection{Direct search bounds}
%%%%%%%%%%%%%%%%%%%%%%

\begin{figure}[t]
    \centering
    \includegraphics[width=0.45\textwidth]{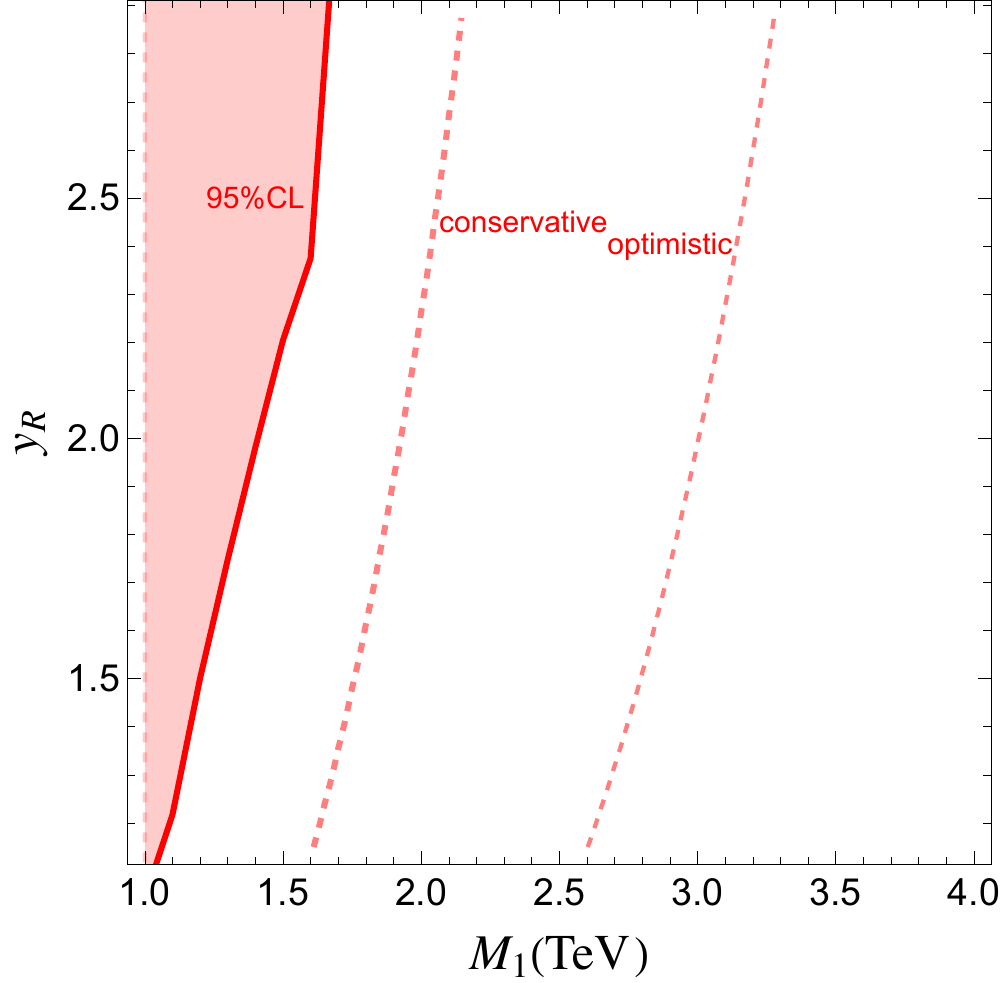} \hspace{0.5cm}
    \includegraphics[width=0.45\textwidth]{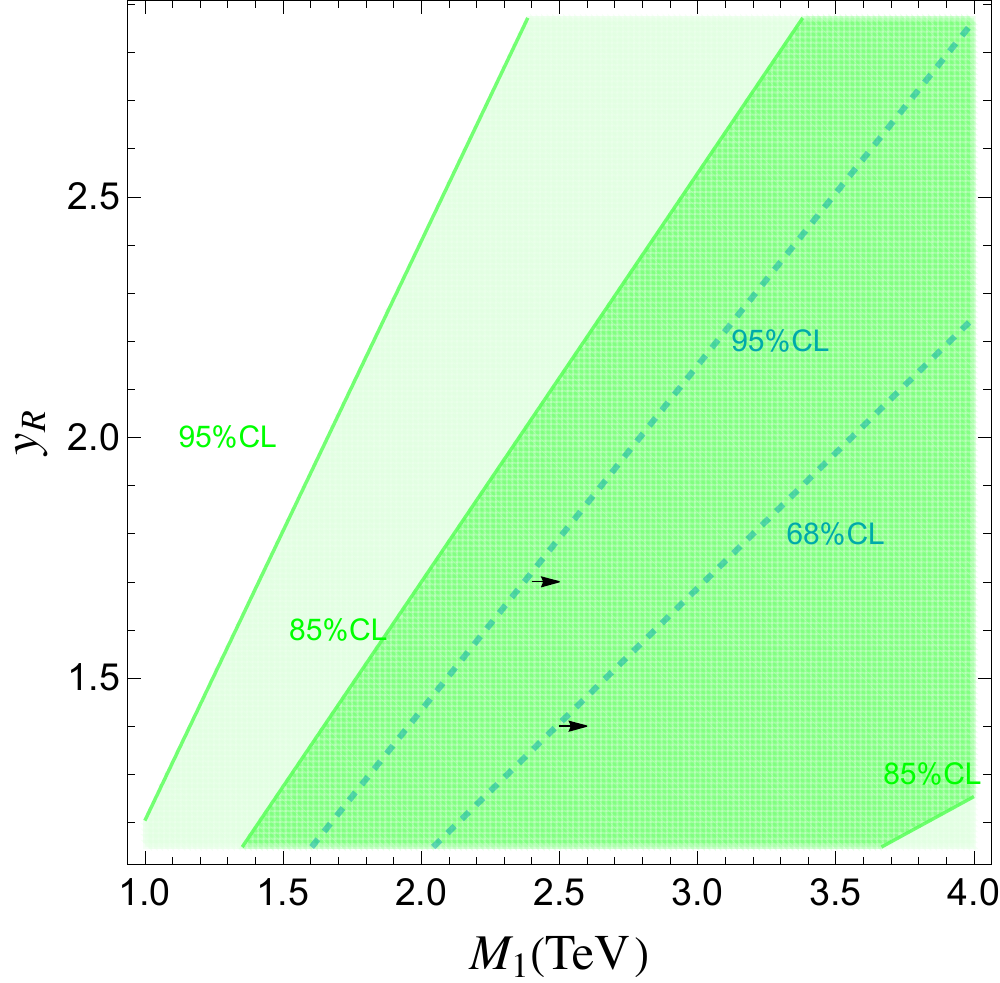} 
    \caption{    
    {\bf Left:} 
    Direct search bounds from single productions of the charge $4/3$ quark in the $M_1-y_R$ plane.   The light red area is excluded at 95\%\,CL by the CMS analysis~\cite{CMS:2017fpk} with an integrated luminosity of $138\,\rm fb^{-1}$.  The two dotted lines are the projected 95\%\,CL reach at HL-LHC with different assumptions on the background.  See text for more details.  
    {\bf Right:} Higgs coupling measurements bounds in the $M_1-y_R$ plane with $M_2$ fixed at $4\,{\rm TeV}$ and $y_L$ marginalized. 
    The green (light green) areas are the region preferred by the current ATLAS Higgs couplings measurement~\cite{ATLAS:2022vkf} at 85\% (95\%\,CL).  Note that, due to a small tension between the BM Model prediction ($\delta \kappa_g>0$) the current Higgs measurement (which prefers $\delta \kappa_g<0$), the lowest CL (with a $\Delta\chi^2$ measured from the $\chi^2_{\rm min}$ of the 3-parameter Higgs coupling fit) is already at $83\%$.  The area right to the cyan dotted lines represent the preferred region of future Higgs measurement at HL-LHC (assumed to be SM-like) with 68\% and 95\%\,CL as labeled.  }
    \label{fig:lhc}
\end{figure}

In the BM model, QCD pair production has the largest rate in the small quark masses region, while single production of mirror quarks dominates for larger masses. 
In the multi-TeV region we are interested in, it turns out that the single production of the charge $-4/3$ quark $X_R$ provides better reach, and we will focus on this channel.  The relevant terms in the Lagrangian in the mass eigenstates are 
\begin{equation}
    -\frac{gs_{R}}{\sqrt{2}}\overline{X}_{R} \gamma ^{\mu } W_{\mu }^{-} b_{R} + {\rm h.c.}\,, 
\end{equation}
where $s_R = \frac{y_R v }{\sqrt{ y_R^2 v^2 +2 M_1^2}}$
is the sine of the mixing angle between the right-handed quarks $b_R$ and $B_R$. 
$X$ decays dominantly to $b W^-$ (or $\bar{b}W^+$), which is very similar to the decay of a charge $2/3$ top partner but with the opposite charge for the $W$. 
In the CMS study~\cite{CMS:2017fpk}, the search on a charge $-4/3$ quark in the single production channel was done with a center-of-mass energy $\sqrt{s}=13\,$TeV and an integrated luminosity of $2.3\infb$.  A lower mass limit of $1.0$\,TeV is obtained assuming a coupling of $0.5$ and 100\% branching ratio to $bW$.   The number of background events and the limits on signal $\sigma \times B(bW)$ are also provided in Ref.~\cite{CMS:2017fpk} up to $M_X = 1.8\,$TeV, making it straightforward to extrapolate the bounds to the BM model parameter space and to the current integrated luminosity, at least up to around $M_1 \approx 1.8$\,TeV.  We use the  VLQ model \cite{Buchkremer_2013} in MadGraph\,5~\cite{Alwall:2014hca} to estimate the number of signal events for a benchmark point in the BM model. 
The resultant exclusion region is shown by the red shaded area in the left panel of \autoref{fig:lhc}, which assumes an integrated luminosity of $138\infb$.  
Furthermore, we extrapolate the bounds to the HL-LHC with an integrated luminosity of $3\inab$.  This goes above the mass range covered in Ref.~\cite{CMS:2017fpk}, and an accurate estimation of the background events is beyond the scope of this paper.  Instead, we consider two scenarios with different assumptions on backgrounds.  The first is simply no background, which gives the best possible reach; for the second, we assume a 400\,GeV invariant-mass window is applied around $m_X$ to efficiently remove backgrounds while keeping most signal events, and then simply extrapolate from the background in the range 1600\,GeV to 2000\,GeV in Ref.~\cite{CMS:2017fpk} to higher invariant masses, assuming the background cross section remains the same. 
This overestimates the background and will thus provide a conservative projection. Both results are shown with red dashed lines in the left panel of \autoref{fig:lhc}, with labels ``optimistic'' and ``conservative'', respectively.       

%

%%%%%%%%%%%%%%%%%%%%%%
\subsection{Higgs coupling measurements}
%%%%%%%%%%%%%%%%%%%%%%

As mentioned in \autoref{sec:bm}, the BM model mainly contributes to $\kappa_b$, $\kappa_g$ and $\kappa_\gamma$ in the Higgs measurements.  Since these couplings contribute to multiple channels as well as the Higgs total width, we perform a 3-parameter global fit to all the Higgs measurements to extract their bounds, assuming all other Higgs couplings are SM-like.  For the current measurements, we use the ones collected in the ATLAS report~\cite{ATLAS:2022vkf}.  Similar reaches are obtained by using  the Higgs measurements at CMS~\cite{CMS:2022dwd}.  
For the HL-LHC Higgs measurements, we use the projections summarized in \cite{deBlas:2022ofj}.  The results of the 3-parameter fit are presented in \autoref{table:hcoupling}. Note that the HL-LHC Higgs measurements are assumed to be SM-like, so that the corresponding central values of $\delta\kappa_b$, $\delta\kappa_g$, $\delta\kappa_\gamma$ are zero by construction.  

\begin{table}[h!]
\centering
\begin{tabular}{|c||c|ccc||c|ccc|} \hline
 & \multicolumn{4}{|c||}{current (ATLAS)} & \multicolumn{4}{c|}{HL-LHC} \\ \hline
 & $1\sigma$ bound &  \multicolumn{3}{c||}{correlation} &  $1\sigma$ bound &  \multicolumn{3}{c|}{correlation} \\
 &  & $\delta\kappa_b$ & $\delta\kappa_g$ & $\delta\kappa_{\gamma}$ & &  $\delta\kappa_b$ & $\delta\kappa_g$ & $\delta\kappa_{\gamma}$  \\ \hline
$\delta\kappa_b$ & $-0.14\pm 0.07$ & 1   &        &   & $\pm $0.0165 & 1   &        &                                   \\
$\delta\kappa_g$ & $-0.069\pm0.05$ & 0.679 & 1 &     &  $\pm $0.012 & 0.725 & 1 &                            \\
$\delta\kappa_{\gamma}$ & $-0.014\pm 0.05$  & 0.323 & $-0.128$ & 1  &  $\pm $0.0125  & 0.313 & $-0.064$ & 1   \\ \hline           
\end{tabular}
\caption{Best fit values $\pm 1 \sigma$ of and correlations among $\kappa_b,\, \kappa_g,\,\kappa_{\gamma}$ from the current ATLAS measurement~\cite{ATLAS:2022vkf} and the HL-LHC projections~\cite{deBlas:2022ofj}, assuming all other Higgs couplings are SM-like.  
}
\label{table:hcoupling}
\end{table}

With \autoref{eq:deltakappa} it is straightforward to map the Higgs coupling constraints in \autoref{table:hcoupling} to the BM model parameter space. 
In particular, we note that the BM model predicts negative $\delta \kappa_b$, $\delta\kappa_\gamma$, and a positive $\kappa_g$.  
Again, we fix $M_2=4\,$TeV, marginalize $y_L$ to project the bounds on the $(M_1,\,y_R)$ plane, which are shown in the right panel of \autoref{fig:lhc}.  The preferred region from the current measurements is shown with light green shades.   
Note that, the current ATLAS measurement prefers a negative $\delta \kappa_g$, which has some tension with the prediction of the BM model.  As a result, the lowered CL in the $(M_1,\,y_R)$ plane is already at 83\% (with $\Delta \chi^2 =3.55$ with respect to the $\chi^2_{\rm min}$ of the 3-parameter Higgs coupling fit), while %certain regions 
a region in the upper-left corner can be excluded by 95\%\,CL.  (We have also shown the $85\%$\,CL contours.)  On the other hand, assuming SM-like results, the HL-LHC prefers the bottom-right region with small $y_R$ and large $M_1$, and the corresponding $68\%$ and $95\%$ CL bounds are shown with cyan dotted lines.    

%

%%%%%%%%%%%%%%%%%%%%%%%%%%%%%%%%%%%%%%%%%%%%%%%%%%%%%%%%%%
\section{Combined Results}
\label{sec:result}
%%%%%%%%%%%%%%%%%%%%%% 

\begin{figure}[h!]
    \centering
    \includegraphics[width=12cm]{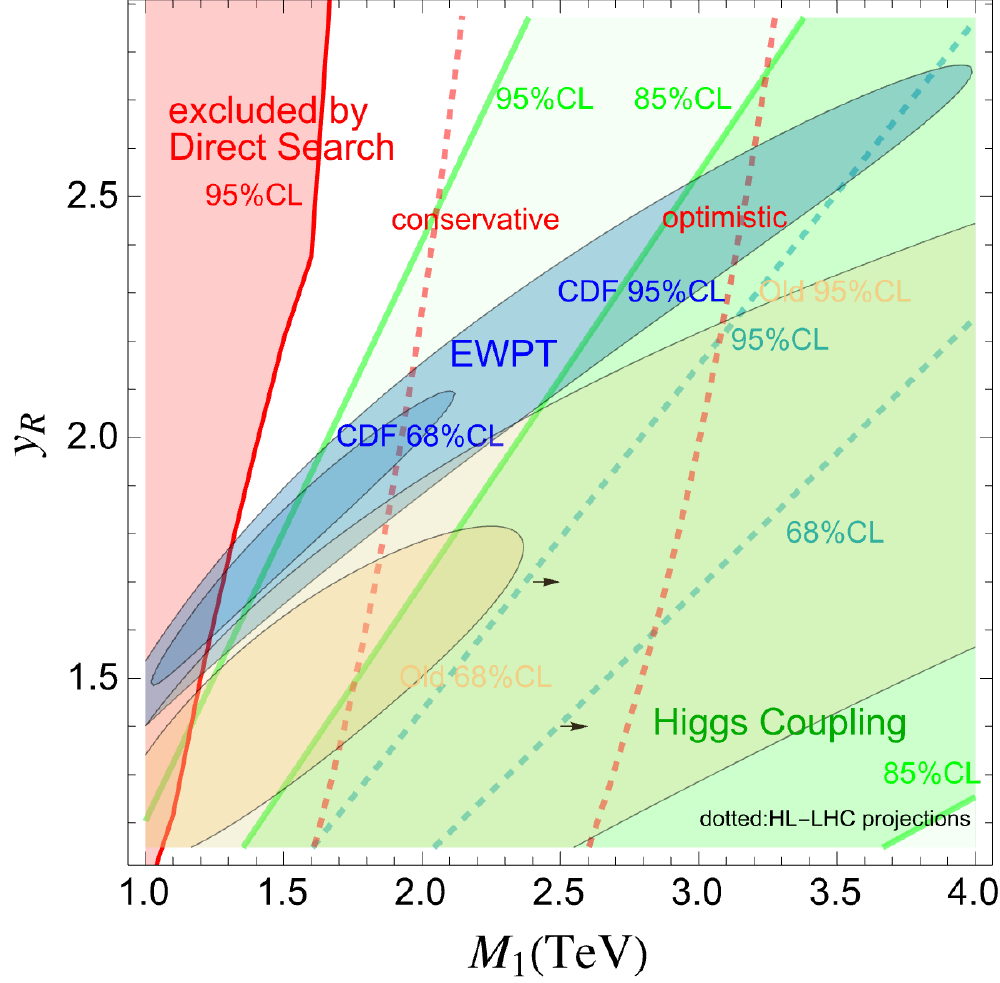}
    \caption{
    Combined results in $M_1-y_R$ plane from the current EW global fit with the new CDF{/old PDG} $m_W$ measurement (preferred regions shown by 
 the blue{/orange} area, from \autoref{fig:bmfit1}), current LHC direct search bounds (excluded region shown by the red area, from \autoref{fig:lhc}), and current LHC Higgs coupling measurements (preferred regions shown by the green areas, from \autoref{fig:lhc}).  The red dotted lines are projected direct-search reaches of the HL-LHC.  The area right to the cyan dotted lines are the preferred regions from the Higgs measurements at HL-LHC, assuming they are SM-like.  }
    \label{fig:combined}
\end{figure}

We now combine the results of the EW precision measurements in \autoref{fig:bmfit1} with the bounds from direct searches and Higgs measurements in \autoref{fig:lhc}.  Our final result is presented in \autoref{fig:combined}.
The main message of our result is that the region in the BM model parameter space preferred by the $\Afb{b}$ and CDF-II $m_W$ measurements 
is still consistent with the current direct search bounds and Higgs measurement constraints at the LHC.  However, the model exhibits some small tension with the Higgs measurements, as it predicts an enhancement on the $hgg$ coupling which is disfavored by current measurements.  %Still, the tension is not statistically significant 
The direct search limit can be significantly improved at the HL-LHC, but even with the most optimistic assumption (with zero background), it will not be able to completely cover the relevant parameter space ($95\%$\,CL preferred region of current EW measurements).  The Higgs couplings measurements provide slightly better reaches.  The region preferred by the HL-LHC Higgs couplings measurements, if SM-like, has little overlap with the one preferred by current EW measurements.  Overall, the possibility that the BM model explains the $\Afb{b}$ and CDF-II $m_W$ discrepancies will likely be either confirmed or ruled out by the HL-LHC.  
{Crucially, this is not the case with the old $m_W$ measurements, for which the (95\%CL) allowed region by the EW measurements extends to much larger values of $M_1$.  The CDF-II $m_W$ result thus provides a strong motivation to search for the BM model at the HL-LHC.}

%%%%%%%%%%%%%%%%%%%%%%%%%%%%%%%%%%%%%%%%%%%%%%%%%%%%%%%%%%
\section{Conclusion}
\label{sec:con}
%%%%%%%%%%%%%%%%%%%%%%

The $m_W$ measurement at CDF II and the $\Afb{b}$ measurement at LEP both exhibit significant discrepancies with the SM predictions.  
Needless to say, further investigations on the experimental side are needed for both measurements before one could make more definitive statements on the origins of the discrepancies.  Meanwhile, the possibility that both discrepancies could come from the same underlying new physics is also interesting and worth exploring.  
In this paper, we consider the Beautiful Mirror model as a possible explanation to both discrepancies. 
By performing a global electroweak fit that includes the $\Afb{b}$ and the CDF-II $m_W$ measurements,  we find that this model  indeed provides a much better fit to the measurements compared with SM.  
To achieve it, the mass of the exotic quark (with charge $-4/3$) is required to be below 4\,TeV at the 95\% confidence level, and the best-fit point corresponds to  a mass of around 1.5\,TeV.   
While the model is consistent with the current direct-search limits at the LHC, the future LHC runs, especially the HL-LHC, will be able to cover most of the regions of the parameter space preferred by the electroweak fit. 
The contributions to the Higgs couplings in this model are also relevant, especially for the $hgg$, $hb\bar{b}$, and $h\gamma\gamma$ couplings.  Once again, the preferred region is consistent with the current LHC measurements but would be in tension with the projected precisions of the Higgs measurement at the HL-LHC, if they turn out to be SM-like.   In summary,  the possibility that both the $\Afb{b}$ and the CDF-II $m_W$ discrepancies are explained by the Beautiful Mirror model will very likely be either confirmed or ruled out after the HL-LHC runs.  

It should also be noted that, while both $m_W$ and observables similar to $\Afb{b}$ could be measured at the LHC~\cite{ATLAS:2017rzl, LHCb:2021bjt, Murphy:2015cha, Yan:2021veo, Dong:2022ayy} or the Electron-Ion Collider (EIC)~\cite{Yan:2021htf, Li:2021uww}, these measurements are difficult and even with the future runs they may not reach the desired precision to resolve the discrepancies.  Future lepton colliders, especially those with Z-pole and $WW$ threshold runs (such as CEPC~\cite{CEPCPhysicsStudyGroup:2022uwl} and FCC-ee~\cite{Bernardi:2022hny}), will be able to significantly improve the measurement precisions of these two observables, as well as the ones of other EW or Higgs measurements.  Such a collider will be able to unambiguously resolve the current observed  discrepancies in $\Afb{b}$ and $m_W$.  

%%%%%%%%%%%%%%%%%%%%%%%%%%%%%%%%%%%%%%%%%%%%%%%%%%%%%%%%%%%%%%%%%%
\subsection*{Acknowledgments}
We thank Henning Bahl for the useful discussions. JG is supported by the National Natural Science Foundation of China (NSFC) under grant No.~12035008.  LTW is supported by the DOE grant DE-SC0013642.

%%%%%%%%%%%%%%%%%%%%%%%%%%%%%%%%%%%%%%%%%%%%%%%%%%%%%%%%%%%%%%%%%%

\bibliographystyle{JHEP}
\bibliography{mw}

\end{document}